\documentclass[superscriptaddress,twocolumn,showpacs,prb]{revtex4-1}
\usepackage{graphicx}% Include figure files
\usepackage{dcolumn}% Align table columns on decimal point
\usepackage{bm}% bold math
\usepackage{color}

\usepackage{subfigure}
\usepackage{graphics} 
\usepackage{graphicx}

\begin{document}

\def\o{\omega}
\def\e{\varepsilon}

\title{Electron energy and temperature relaxation in graphene on a piezoelectric substrate}
\author{S. H. Zhang}
\affiliation{Department of Physics, University of Antwerp, Groenenborgerlaan 171, B-2020 Antwerpen, Belgium}
\affiliation{Key Laboratory of Materials Physics, Institute of Solid State Physics, Chinese Academy of Sciences, Hefei 230031, China}
\affiliation{Beijing Computational Science Research Center, Beijing 100084, China}
\author{W. Xu}
\affiliation{Key Laboratory of Materials Physics, Institute of Solid State Physics, Chinese Academy of Sciences, Hefei 230031, China}
\author{F. M. Peeters}
\affiliation{Department of Physics, University of Antwerp, Groenenborgerlaan 171, B-2020 Antwerpen, Belgium}
\author{S. M. Badalyan}
\email{samvel.badalyan@uantwerp.be}
\affiliation{Department of Physics, University of Antwerp, Groenenborgerlaan 171, B-2020 Antwerpen, Belgium}
\affiliation{Department of Physics and Astronomy, University of Missouri-Columbia, Missouri 65211, USA}

\begin{abstract}
We study the energy and temperature relaxation of electrons in graphene on a piezoelectric substrate. Scattering from the combined potential of extrinsic piezoelectric surface acoustical (PA) phonons of the substrate and intrinsic deformation acoustical (DA) phonons of graphene is considered for a (non-)degenerate gas of Dirac fermions. It is shown that in the regime of low energies or temperatures the PA phonons dominate the relaxation and change qualitatively its character. This prediction is relevant for quantum metrology and electronic applications using graphene devices and suggests an experimental setup for probing electron-phonon coupling in graphene.
\end{abstract}

\pacs{72.80.Vp, 72.10.-d, 78.67.Wj, 72.20.Jv}

\maketitle

\section{Introduction}

Graphene \cite{NovGeim2004,NovGeim2005} is a promising material for future applications in electronic and optical technologies \cite{GeimAMD2007,Castro2009} and for quantum metrology \cite{Woszczyna2012,Falko2013}. For anticipated new functionalities, electron-phonon scattering in graphene structures \cite{Ando} can play a crucial role. It can drastically affect the operating performance of  information processing devices and the harvesting efficiency of photoelectrons in optoelectronic devices. It limits also the breakdown current of quantum correlations in graphene. Therefore, a rigorous understanding of the energy coupling mechanisms between Dirac fermions in graphene and lattice vibrations is not only of fundamental interest but also of great practical relevance.
Several theoretical \cite{Song,Bistritzer,Tse,Kubakaddi,Viljas} and experimental \cite{Baker1,Webber,Baker2,Fong1,Fong2,Betz1,Betz2,Graham,Tan} works have addressed this interesting key problem in a broad range of temperatures down to the sub-Kelvin regime \cite{Borzenets} with significant discrepancies found both in theoretical results \cite{Tse,Kubakaddi} and measurements \cite{Baker1,Tan,Betz1}. Most of graphene devices studied so far are deposited on a SiO$_{2}$ substrate. According to recent theoretical predictions \cite{Chen},  disorder effects in such structures can substantially modify the results obtained in the clean limit \cite{Bistritzer,Tse,Kubakaddi,Viljas}. 

Recently graphene devices on top of a GaAs substrate were fabricated \cite{Woszczyna2012,Woszczyna2011}. GaAs has a much better surface quality, its stronger hydrophilicity prevents folding of large-scale graphene flakes while its substantially larger dielectric constant improves the screening of substrate defects. Such high-purity GaAs-supported graphene device structures allow high precision measurements and can provide an alternative reliable source of information on electron-phonon coupling in graphene.

In this paper we study the electron energy and temperature relaxation in a degenerate and non-degenerate gas of massless Dirac fermions in monolayer graphene supported by a piezoelectric substrate. We focus mainly on the low temperature regime \cite{Borzenets,Betz1,Baker1,Baker2} where surface optical phonon modes of the substrate and graphene \cite{Fratini2008,Perebeinos2009,Avouris} as well as the disorder-assisted and two-phonon scattering mechanisms \cite{Song} are not effective in carrier relaxation processes. 
Recently other remote phonon scattering mechanisms have been considered that are due to hybrid interface modes \cite{Perebeinos20099,Ong2013}. In the low temperature regime Rayleigh-like surface acoustical phonons \cite{Landau} play an important role in carrier scattering \cite{Thalmeier2010,Zhang2013,SMB1988,SMB1989,SMB1991}.  Piezoelectric surface acoustic waves can be generated and manipulated electrically and, as demonstrated recently, open new possibilities in highly promising directions of quantum phononics \cite{Clerk2012,Gustafsson2012} and plasmonics \cite{Farhat2013,Schiefele2013}.

Here we calculate carrier relaxation through scattering from the potential of extrinsic piezoelectric surface acoustical (PA) phonons of the substrate and from the potential of intrinsic deformation acoustical (DA) phonons of graphene. We derive analytical formulae for the relaxation rates in the energy and temperature regions where PA and DA phonon scattering is qualitatively different. It is shown that the characteristic electron energy and electron temperature relaxation times, $\tilde{\tau}(\e)$ and $\tilde{\tau}(T_{e})$, are independent of the carrier density both in a Boltzmann gas and in the Bloch-Gr\"unison (BG) regime of a Fermi gas. In both these regimes  $\tilde{\tau}(\e)$ exhibits an inverse linear (quadratic) dependence for the PA (DA) phonons with the electron energy $\e$ and the lattice temperature $T$, respectively. Similar dependences we find for the PA and DA phonon contributions to $\tilde{\tau}(T_{e})$ versus the electron temperature $T_{e}$, independently of the degree of the degeneracy of the electron gas. 
Thus, this universal weakening ({\it linear versus quadratic}) of the energy and temperature dependence of the relaxation rates for PA versus DA phonon scattering makes the PA phonon relaxation mechanism with {\it small angle scattering events} dominant in the regime of low energies and temperatures. 
This prediction suggests a new experimental approach for studying electron-phonon relaxation in graphene and its control via extrinsic piezoelectric coupling.

The paper is organized as follows. In Sec.~II we represent electron-phonon interaction vertices for the PA and DA phonons and in Sec.~III discuss the electron energy relaxation in the (non)degenerate regimes of Dirac fermions. Then, in Sec.~IV we present our results for the electron temperature relaxation in the Fermi and Boltzmann gas of Dirac fermions with a detailed comparison of the respective relaxation rates for the PA and DA phonons. In Sec. V we conclude with a summary of the behavior of the PA and DA phonon relaxation rates calculated, in typical scattering regions.

\section{Interaction vertices for PA and DA phonon scattering} 

The graphene sheet is laid on top of the (100) surface of GaAs, which is spanned by the $x$ and $y$ lattice axes. In substrates with lack of center of symmetry such as GaAs, the elastic displacement of surface acoustic waves induces a piezoelectric polarization of the lattice, which leads to an electric potential \cite{Levinson1996,Zhang2013} both inside and outside the crystal that couples to Dirac fermions in graphene. Similar to the previous study on conventional two-dimensional gases \cite{Levinson1996}, the electron interaction vertex in graphene for the PA phonon with a wave vector ${\bf q}$ can be written as
\begin{equation}\label{PA}
\left|\gamma^{PA}_{q}\right|^{2}=\frac{c_{PA}^{2} \hbar^{2}  \left(\hat{q}_{x}\hat{q}_{y}\right)^{2}v_{PA}e^{-2 q d}}{p_{0}{\bar\tau}_{PA}}~.
\end{equation}
Here $\hat{q}_{x,y}\equiv q_{x,y}/q$, $1/{\bar \tau}_{PA}=(e\beta)^2p_0/(2\pi\hbar\rho_s v_{PA}^2)$ with a nominal piezoelectric scattering time ${\bar \tau}_{PA}\approx 8$ ps, and $\rho_s=5.3$ g/cm$^{3}$ and $e\beta=2.4\times 10^7$ eV/cm corresponding to the mass density and the single modulus of the piezoelectric tensor of the GaAs cubic crystal, and $p_0=2.5\times10^6$ cm$^{-1}$ a characteristic wave vector scale in GaAs \cite{GL}. 
The numerical factor $c_{PA}\approx 4.9$ and the velocity of surface acoustical waves, $v_{PA}\approx 2.7 \times10^3$ m/s, are determined by elasticity constants of GaAs. Notice that in contrast to the vertex of electron interaction with bulk piezoelectric acoustical phonons, the vertex $\gamma_{q}^{PA}$ for the surface PA phonons does not depend on the magnitude of $\vec{q}$ but only on its orientation. For typical distances $d\sim 5$ \AA~  between the substrate and graphene \cite{Woszczyna2012,Woszczyna2011}, one can approximate $e^{-2 q d}\approx 1$ so that  taking $\left(\hat{q}_{x} \hat{q}_{y}\right)^{2}\approx 1/4$, the PA vertex $\gamma_{q}^{PA}$ becomes independent of $q$ for the strongest interaction along the diagonal $q_{x}\approx q_{y}$. 

This contrasts the linear $q$ dependence of the electron interaction vertex for the intrinsic deformation potential in graphene \cite{Bistritzer,Tse,Kubakaddi,Viljas}. We write the DA vertex 
%$\left|\gamma_q^{DA}\right |^{2}=\hbar^2 q v_{DA} / \left(p_0^2 \bar{\tau}_{DA}\right)$ 
\begin{equation}\label{DA}
\left|\gamma_q^{DA}\right |^{2}=\frac{\hbar^2 q v_{DA}}{p_0^2 \bar{\tau}_{DA}}
\end{equation}
introducing a nominal scattering time $1/\bar{\tau}_{DA}= \Xi^{2} p_0^2 / \left(2 \hbar \rho_{gr} v^{2}_{DA}\right)$ for DA phonons. We obtain $\bar{\tau}_{DA}\approx 0.8$ ps using the following values for graphene parameters \cite{Kaasbjerg2012,Avouris}: $\Xi=6.8$ eV the coupling constant of the deformation potential, $\rho_{gr}=7.6 \times 10^{-7}$ kg m$^{-2}$ the surface mass density of graphene, and $v_{DA}=2.0\times10^4$ m/s the sound velocity in graphene. The ratio $\left|\gamma_q^{PA}\right|^2 / \left|\gamma^{DA}_{q}\right|^{2}\approx 2.0\times10^{5}~\text{cm}^{-1}/q$ suggests that depending on the carrier density, the energy and temperature, both the PA and DA phonons can provide important channels for carrier relaxation in graphene.

Notice that the deformation potential interaction strength in graphene has not a well established value yet \cite{Fischetti2013}. Strengths ranging from small $\Xi=3.2$ eV \cite{Borisenko2010} to large values $\Xi=20$ eV \cite{Suzuura2002} have been used in the literature to describe experiment. We take an intermediate value of $\Xi=6.8$ eV that was obtained from a first-principles DFT approach \cite{Kaasbjerg2012}. We emphasize, however, that our qualitative conclusions on the dominance of the PA phonon relaxation mechanism at low temperatures are not affected by the choice of the value of $\Xi$. It determines only the transition temperature of the PA mechanism dominance, which  remains measurable for even larger values of $\Xi$ in the sub-Kelvin relaxation regime \cite{Borzenets}.

We emphasize also that the deformation potential constant $\Xi$ in Eq.~(\ref{DA}) obtained in phonon scattering experiments \cite{Chen2008,Efetov2010} includes the static dielectric screening effect of the surrounding environment of the individual graphene sheet. Therefore, one can expect that the value of $\Xi$ in the GaAs supported graphene should be smaller than that in SiO$_2$ supported graphene by a factor of $(1+\kappa_{\text{GaAs}})/(1+\kappa_{\text{SiO}_2}) \approx 2.9$ \cite{Landau2,SMB2012} where $\kappa_{\text{SiO}_{2}}=3.8$ \cite{Kim2011} and $\kappa_{\text{GaAs}}=12.9$ \cite{GL} are respectively the low frequency dielectric constant of SiO$_{2}$ and GaAs. This is another important argument justifying our choice of the actual value for the constant $\Xi=6.8$ eV. % in comparison with $\Xi~20$ eV reported in previous works \cite{} for SiO$_2$. 
As to the environmental screening of piezoelectric interaction, the PA vertex in Eq.~(\ref{PA}) includes the background dielectric constant \cite{Levinson1996}, which is taken into account in the calculation of the constant $c_{PA}$.

\section{Electron energy-loss power} 

The energy-loss power of a Dirac fermion due to scattering from the $s=$ PA, DA phonon field
\begin{equation}\label{Qea}
Q^s(\e_{\bf \lambda k}) \equiv \frac{\e_{\bf \lambda k}-\e_{s}^{*}}{\tilde{\tau}^{s}(\e_{\bf \lambda k})} = Q^{s}_{+}(\e_{\bf \lambda k}) - Q^{s}_{-}(\e_{\bf \lambda k})
\end{equation}
where $\tilde{\tau}^{s}(\e_{\bf \lambda k})$ is the characteristic electron energy relaxation time. 
%The critical energy $\e_{s}^{*}$, at which the electron energy-loss power changes its sign, depends on the degeneracy regime of the electron gas and the electron-phonon scattering mechanism \cite{GL}. 
In the electron-phonon scattering process only electrons with sufficient high energy can lose energy by emitting phonons, otherwise they can only gain energy by absorbing phonons. The critical energy $\e_{s}^{*}$ in Eq. (\ref{Qea}) provides a reference point for the energy, above or below which an electron gains or loses its energy. This critical energy depends on the degeneracy regime of the electron gas and the electron-phonon scattering mechanism \cite{GL}. 
The functions $Q^{s}_{\pm}(\e_{\bf \lambda k})$ are defined as 
\begin{equation}
Q^{s}_{\pm}(\e_{\bf \lambda k})=\sum_{\lambda' {\bf k}' {\bf q}}(\e_{\bf \lambda k}-\e_{\bf \lambda' k'}) W^{\pm s {\bf
q}}_{\lambda {\bf k}\rightarrow \lambda' {\bf k}'} {1-f_{T}(\e_{\lambda' \bf k'})\over1-f_{T}(\e_{\lambda \bf k})}
\end{equation}
with $\pm$ signs corresponding to the phonon emission ($+$) and absorption ($-$) processes. 
In monolayer graphene the electron energy disperses linearly with its momentum ${\bf k}$, $\e_{\lambda \bf k}=\lambda v_{F} k$, where $\lambda$ and $v_{F}$ are the electron chirality and Dirac band velocity, respectively. The Fermi function $f_{T}(\e_{\lambda \bf k})$ determines the carrier distribution for given Fermi energy, $\e_{F}$, and temperature, $T$. 
The electron transition probabilities are 
\begin{eqnarray}
W^{\pm s {\bf q}}_{\lambda {\bf k}\rightarrow \lambda' {\bf k}'}&=&{2\pi\over\hbar} \left| M^{\pm s {\bf q}}_{\lambda {\bf k}\rightarrow \lambda' {\bf k}'}\right|^{2} \left( N_{T}\left(\o_{s {\bf q}}\right) + \frac{1}{2} \pm \frac{1}{2} \right) 
\\ \nonumber 
&\times&
\delta\left(\varepsilon_{\lambda \bf k}-\varepsilon_{\lambda' \bf k'} \mp \hbar \omega_{s {\bf q}}\right)~.
\end{eqnarray}
The Bose factor $N_{T}\left(\o_{s {\bf q}}\right)$ gives the number of $s {\bf q}$ phonon modes with energy $\o_{s {\bf q}}=v_{s}q$ at the lattice temperature $T$. Using the wave functions of Dirac fermions, $\psi({\bf R})=\left( e^{-i\theta_{\bf k}},\lambda \right )^{T} e^{-i {\bf k}\cdot{\bf R}}/\sqrt{2\cal{A}}$, where $\theta_{\bf k}$ is the polar angle of the vector ${\bf k}$ and $\cal{A}$ the normalization area, we can write 
\begin{eqnarray}
\left|M^{\pm s {\bf q}}_{\lambda {\bf k}\rightarrow \lambda' {\bf k}'}\right|^{2}=\frac{\delta_{\bf k', k \mp q}}{\cal{A}} \left|\gamma^{s}_{q}\right|^{2} {\cal F}_{\lambda'\lambda}(\theta_{{\bf k}{\bf k}'})~.
\end{eqnarray}
Here, $\theta_{\bf k k'}=\theta_{\bf k'}- \theta_{\bf k}$ and the form factor ${\cal F}_{\lambda' \lambda}(\theta_{\bf k
k'})=\left(1+\lambda\lambda'\cos \theta_{\bf k k'} \right)/2$ represents an overlap of the spinors. % wave functions.

\subsection{Energy relaxation in a Fermi gas of Dirac fermions}

In doped graphene samples scattering from acoustical phonons in a Fermi gas of Dirac fermions is always kinematically quasielastic, $|\e-\e_{F}|, T \ll \e_{F}$, and only intra-chirality subband electronic transitions contribute to the energy relaxation. Taking $\lambda=\lambda'=1$, we can represent the energy-loss power due to extrinsic PA and intrinsic DA phonons as
\begin{equation}\label{eel}
Q^{s}(\e_{\bf k})={a_s c_s^2\over\pi} \frac{\hbar v_{s} k}{{\bar \tau}_{s}} \left({k \over p_0}\right)^{1+m}{\cal{G}}_{1+m}(a_s,x)~.
\end{equation}
Here we introduce the function ${\cal G}_{m}(a_s,x)={\cal G}^{+}_{m}(a_s,x) - {\cal G}^{-}_{m}(a_s,x)$ with
\begin{equation}\label{calg}
{\cal G}^{\pm}_m(a_{s}, x)=\int_0^{z^{s}_\pm} dz z^m \eta^{\pm}(a_{s}, z) \psi^{\pm}(x,y)
\end{equation}
and $z=q/2k$ and $z^{s}_{\pm}=1/(1\pm  a_{s})$. In Eq.~(\ref{eel}) we take for the PA phonons $m = 0$ and $a_s = a_{PA} = v_{PA}/v_F$, $c_s = c_{PA}=4.9$ and for the DA phonons $m = 1$, $a_s=a_{DA}=v_{DA}/v_F$, $c_s=c_{DA} = 2\sqrt{2}$. For brevity, we omit the chirality indices and introduce also the functions $\psi^{\pm}(x,y)=(N(y)+1/2\pm1/2)(1-f(x\mp y))/(1-f(x))$ where $x=(\e_{k}-\varepsilon_F)/T$ and $y=\hbar\o_{s q}/T= z (k/k_F)(T_{BG}^{s}/T)$ and the characteristic Bloch-Gr\"uneisen (BG) temperature is $T^{s}_{BG}=2\hbar v_{s} k_{F}$. For the PA and DA phonons, respectively, $T^{s}_{BG}\approx 7.3 \sqrt{n}$ K and $\approx 54 \sqrt{n}$ K (here the electron density $n$ is taken in units of $10^{12}$ cm$^{-2}$). The functions $\eta^{\pm}(a_{s},z)=\sqrt{1-( a_{s}\pm(1-a_{s}^2)z)^2}/(1-a_{s}^2)$ describe the overlapping of the spinor wave functions. Because the Fermi velocity in graphene $v_F\approx 1.0\times 10^6$ m/s is much larger than the sound velocity in GaAs and in graphene, in our analytical calculations we use the approximation $\eta(a_{s}, z)\approx \eta(0,z)=\sqrt{1-z^2}$ and $z^{s}_{\pm}\approx1$.

In the low $T$ regime, $T\ll T^{s}_{BG}$, if the energy of a test electron is small, $\e-\e_{F}\ll T_{BG}^{s}$, electronic transitions are dominated by small angle scattering events. The typical phonon momenta, $q \ll k_{F}$, correspond to the $z\rightarrow 0$ limit and one can take the integration over $z$ in Eq.~(\ref{calg}) up to infinity. Replacing also $1-z^{2}$ by $1$ and taking into account that $z\approx y\left(T/T^{s}_{BG} \right)$, we have ${\cal{G}}_n(x)\approx(T/T^{s}_{BG})^{n+1}{\cal{F}}_n(x)$ where we define ${\cal{F}}_n(x)={\cal F}^{+}_n(x)-{\cal F}^{-}_n(x)$ with the functions ${\cal F}^{\pm}_n(x)=\int_0^{\infty} dy y^n\psi^{\pm}(x,y)$. Hence, the relaxation time in the low $T$ regime is given as
\begin{eqnarray}\label{leelr}
Q^{s}(\e_{\bf k})={a_sc_s^2\over\pi} \left({\e_{F} \over \e_{0} }\right)^{1+m}
\left({T\over T^{s}_{BG}}\right)^{1+m} \frac{\e_{\bf k}-\e_{F}}{{\bar \tau}_{s}}\frac{{\cal F}_{1+m}(x)}{x}~.
\end{eqnarray}
Here the energy scale $\e_{0}=\hbar v_{F} p_{0}\approx 1924$ K. 

For {\it thermal electrons}, $|\e_k-\e_{F}|\lesssim T$, small angle nonelastic scattering events with typical phonon momenta $q\sim T/ v_{s}\ll \hbar k_{F}$ and energies $\o_{s {\bf q}}\sim T$ dominate the relaxation. The functions ${\cal F}_1(x)\approx 2 \log(2) x$ and ${\cal F}_2(x)\approx (\pi^{2}/3) x$ are linear in $x$ for $x\ll 1$ hence the energy-loss power $Q^{s}(\e_{\bf k})$ exhibits a linear and quadratic $T$ dependence, respectively, for the PA and DA phonons. In this regime for both mechanisms the energy relaxation time $\tilde{\tau}^{s}(\e)$ does not depend on the carrier energy and density. The ratio $Q^{PA}/Q^{DA}\approx 9.7 / T  [\text{K}]$ so that at low temperatures the PA phonon contribution to the electron energy relaxation rate dominates.

In the regime of {\it hot electrons}, $T\ll |\e_k-\e_{F}| $, electronic transitions are dominated with spontaneous emission of phonons. We have ${\cal{F}}_n(|x|)\approx |x|^{n+1}/(1+n)$ for $|x|\gg1$ so that the energy-loss power $Q^{s}(\e_{\bf k}) \propto \left(\e_k-\e_F\right)^{2+m}$ is independent of $T$. The comparison gives $Q^{PA}(\e_{\bf k})/Q^{DA}(\e_{\bf k})\approx 34.4/|\e_{\lambda k}-\e_F| [\text{K}]$ so that at low energies the extrinsic PA phonons dominate the energy relaxation also in this subregime.

\subsection{Energy relaxation in a Boltzmann gas of Dirac fermions}

\begin{figure}[t]
\includegraphics[width=\columnwidth]{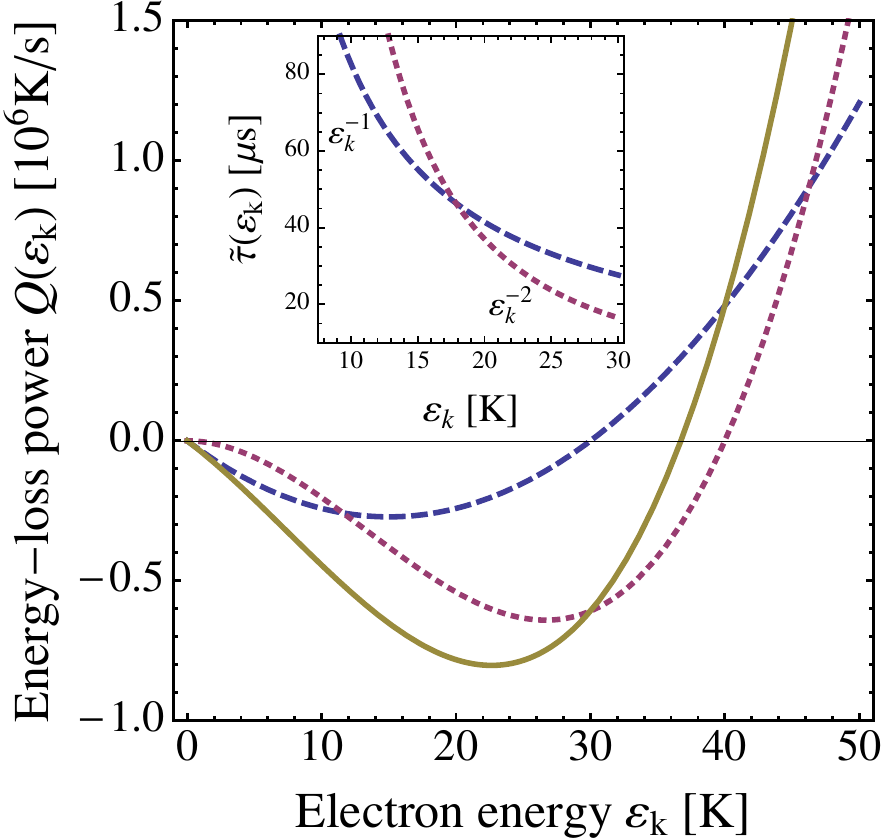}
\caption{Electron energy-loss power in neutral graphene at $T=10$ K. Inset plots the characteristic PA and DA energy relaxation times and shows the suppression of the DA relaxation at low energies in comparison with that through PA phonons. 
The dashed, dotted, and solid curves correspond, respectively, to the PA, DA, and PA+DA combined relaxation mechanisms.}
\label{fig1}
\end{figure}

In almost neutral graphene samples the Pauli principle does not play an important role and we have a Boltzmann gas of Dirac fermions, ($ \e_{F} \ll T $). We represent the energy-loss power as
\begin{equation}\label{eelrB}
Q^{s}(\e_{\bf k})={a^{2}_s c_s^2\over\pi} \left({\e_{\bf k} \over \e_0}\right)^{1+m} \frac{\e_{\bf k}}{{\bar \tau}_{s}}  
{\cal{B}}_{1+m}(a_s,T)
\end{equation}
where the function ${\cal B}_{m}(a_s, T)={\cal B}^{spnt}_{m}(a_s, T) + {\cal B}^{ind}_{m}(a_s, T)$ is introduced. The first term, ${\cal B}^{spnt}_{1+m}(a_{s}, T)=\int_0^{z^{s}_+} dz z^{1+m} \eta^{+}(a_{s}, z)=b_{s}$ with $b_{PA}=1/3$ and $b_{DA}=\pi/16$, is related to the spontaneous phonon emission in the absence of any phonons in the crystal. %Here $z=q/2k$ and $z^{s}_{\pm}=1/(1\pm  a_{s})$, the function $\eta^{\pm}(a_{s},z)=\sqrt{1-(a_{s}\pm(1-a_{s}^2)z)^2}/(1-a_{s}^2)$. 
The other term, ${\cal B}^{ind}_{1+m}(a_{s}, T)={\cal I}^{+}_{1+m}(a_{s}, T)-{\cal I}^{-}_{1+m}(a_{s}, T)$, is related to the energy gain by an electron due to the induced equilibrium phonons. Here ${\cal I}^{\pm}_{1+m}(a_{s}, T)= \int_0^{z^{s}_\pm} dz z^{1+m} \eta^{\pm}(a_{s}, z)N_{T}(y)$ and  $y=2 a_{s} z (\e_{\bf k}/T)$. In graphene $a_{s}\ll 1$ and electron-phonon scattering is quasielastic in a Boltzmann gas. Hence, processes with $y\ll 1$ and a large number of induced phonons, $N(y)\approx 1/y$, dominate the energy relaxation. To leading order in $v_{s}/v_{F}$, we obtain ${\cal B}^{ind}_{1+m}(a_{DA}, T)\approx -d_{s} T/\e_{\bf k}$ with $d_{PA}=1$ and $d_{DA}=\pi/4$, respectively, for $m=0$ and $m=1$. Then, the total energy-loss power is 
\begin{equation}
Q^{s}(\e_{\bf k})=\frac{a^{2}_s b_{s} c_s^2}{ \pi } \left({\e_{\bf k} \over \e_0}\right)^{1+m} \frac{\e_{\bf k} - \e_{s}^{*}}{{\bar \tau}_{s}}~.
\end{equation}
We find that the critical energy $\e_{s}^{*}$, at which the electron energy-loss power changes its sign in a Boltzmann gas, differs for the PA and DA mechanisms, i.e. $\e_{PA}^{*}=3 T$ and $\e_{DA}^{*}= 4 T$ in graphene supported by a piezoelectric substrate. Therefore, an electron with higher energies $\e_{\bf k}$ close to $\e_{DA}^{*}= 4 T$ loses its energy mainly due to scattering from the PA phonons while the DA phonons play a secondary role. At lower energies $\e_{\bf k}$ close to $\e_{PA}^{*}= 3 T$, the electron gains energy from the DA phonons and the relaxation channel due to the PA phonons is suppressed ({\it cf.} Fig.~\ref{fig1}). 
Additionally, the  PA and DA relaxation channels can be identified, respectively, through the inverse linear and quadratic energy dependence of the energy relaxation time, $\tilde{\tau}^{PA}(\e_{\bf k})\propto \e^{-1}_{\bf k}$ and $\tilde{\tau}^{DA}(\e_{\bf k})\propto \e_{\bf k}^{-2}$ ({\it cf.} the inset in Fig.~\ref{fig1}). We obtain the ratio $\tilde{\tau}^{PA}(\e_{\bf k})/\tilde{\tau}^{DA}(\e_{\bf k}) \approx \e_{\bf k}/17.9 \text{ K}$ and at low energies the energy relaxation through the PA phonons is a faster process in nearly neutral graphene.

\section{Electron temperature relaxation}

When a gas of Dirac fermions is driven out of thermodynamic equilibrium under the influence of external forces, ultrafast carrier collision processes set the electronic equilibrium on a time-scale of 30 fs \cite{Sun2008,Dawlaty2008,Breusing2009,Johannsen2013}. The established distribution, $f_{T_{e}}(\e_{\lambda \bf k})$, is described by the well-defined carrier temperature, $T_e >T$, the same for the electron and hole gases. 
The number of thermally excited electrons or holes is 
%
%$n_{e, h}(T_{e})=g T_{e}^{2}F_{1}(\pm \bar{\mu})/2\pi v^{2}_{F}$ 
%
\begin{equation}
n_{e, h}(T_{e})=\frac{g}{2\pi v^{2}_{F}} F_{1}\left(\pm \bar{\mu}(T_{e})\right) T_{e}^{2}
\end{equation}
with $g=4$ the spin and valley degeneracy and the Fermi integrals $F_{n}(\bar{\mu})=\int_{0}^{\infty}  dx x^{n} \left(\exp(x-\bar{\mu})+1\right)^{-1}$ where $\bar{\mu}=\mu(T_{e})/T_{e}$. The chemical potential, $\mu(T_{e})$, is determined from the conservation of particles, $n=n_{e}(T_{e})-n_{h}(T_{e})$, where $n$ is the electron density at $T_{e}=0$.

The electron temperature relaxation in a gas of Dirac fermions is described by the cooling power per electron
%, $\bar{Q}^s(T_e,T)=\bar{Q}_{+}^s(T_e,T)-\bar{Q}_{-}^s(T_e,T)$,
% 
\begin{equation}
\bar{Q}^s(T_e,T)=\bar{Q}_{+}^s(T_e,T)-\bar{Q}_{-}^s(T_e,T)
\end{equation}
where 
\begin{eqnarray}\label{ETP}
\bar{Q}_{\pm}^s={g\over n_{e}(T_{e}) {\cal A}} \sum_{{\bf k}{\bf k}'}^{\bf q} \hbar \omega_{s {\bf q}} W^{\pm s {\bf q}}_{ {\bf k}\rightarrow  {\bf k}'} f_{T_{e}}(\e_{\bf k})(1-f_{T_{e}}(\e_{\bf k'}))~.
\end{eqnarray}
The characteristic electron temperature relaxation time, $\tilde{\tau}^{s}({T_{e}})$, can be expressed in terms of the cooling power from the balance equation
\begin{eqnarray}\label{ETCT}
\tilde{\tau}^{s}(T_{e},T)=\frac{\bar{\e}(T_{e})-\bar{\e}(T)} {\bar{Q}^s(T_e,T)} ~.
\end{eqnarray}
Here we define the average electron energy in graphene as $\bar{\e}(T_{e})=T_{e} F_{2}\left(\bar{\mu}(T_{e})\right)/ F_{1}\left(\bar{\mu}(T_{e})\right)$.

\begin{table*}[t]
\caption{The energy, temperature, and density dependence of the electron energy and temperature relaxation rates.}
\label{tab}
\begin{ruledtabular}
\begin{tabular}{lll|ll}
& \multicolumn{2}{c|}{$Q^{s}(\e_{\bf k})$,~~~$\tilde{\tau}^{s}(\e_{\bf k})$} & \multicolumn{2}{c}{$\bar{Q}^{s}(T_{e},T)$,~~~$\tilde{\tau}^{s}(T_{e})$ ~~~($T \ll T_{e}$)}
\\ \hline
& Fermi gas\footnote{Here results are given for the thermal regime, $|\e-\e_{F}| \lesssim T$. For hot electrons, $T\ll |\e-\e_{F}|$, one should replace $T$ by $|\e-\e_{F}|$ in these formulas and take $\tilde{\tau}^{PA} / \tilde{\tau}^{DA} \approx |\e-\e_{F}| [\text{K}]/34.4$. } ($T \ll T_{BG}^s$) & Boltzmann gas ($\e_F \ll T $) & Fermi gas ($T_{e}\ll T_{BG}^s$) & Boltzmann gas ($\e_F \ll T_{e} $) 
\\ \hline
s=PA & ~~$(\e_{\bf k}-\e_F)T,~~~T^{-1}$ & ~~~$\e_{\bf k}(\e_{\bf k}-3T),~~~\e_{\bf k}^{-1}$ & ~~~~$n^{-1/2}T_{e}^{3},~~~T_{e}^{-1}$ & ~~~~~~$T_{e}^{2},~~~T_{e}^{-1}$ \\ 
s=DA & ~~$(\e_{\bf k}-\e_F)T^2;~~~T^{-2}$ & ~~~$\e_{\bf k}^2(\e_{\bf k}-4T);~~~\e_{\bf k}^{-2}$ & ~~~~$n^{-1/2}T_{e}^{4},~~T_{e}^{-2}$ & ~~~~~~$T_{e}^{3},~~~T_{e}^{-2}$ \\ 
$\tilde{\tau}^{PA} / \tilde{\tau}^{DA}$ & $~~~~~~~~~~T [\text{K}]/9.7$ & $~~~~~~~~~~~\e_{\bf k} [\text{K}]/17.9$ & $~~~~~~~T_{e} [\text{K}]/8.9$ & $~~~~~~T_{e} [\text{K}]/4.3$  \\
\end{tabular}
\end{ruledtabular}
\end{table*}

%\end{widetext}

\subsection{Electron temperature relaxation in a Boltzmann gas of Dirac fermions}

In a Boltzmann gas of Dirac fermions ($ \e_{F} \ll T_{e}$), the chemical potential is approximated by $\mu(T_{e})\approx \e_{F}^{2}/ 4\ln2 T_{e}$  and we find $\bar{\e}(T_{e})-\bar{\e}(T)=18 \zeta (3)/\pi ^2 \left(T_{e}-T\right)$. A direct calculation shows that in the limit of $T_{e}\gg T$, the cooling power is
\begin{eqnarray}\label{eelrB}
\bar{Q}^s(T_e)&=& B_{s}  \left({T_{e} \over \e_0}\right)^{1+m} {T_{e}\over {\bar \tau}_{s}}
\end{eqnarray}
where $B_{s}=\left(a^{2}_s b_{s} c_s^2/ \pi \right) F_{3+m}\left(\bar{\mu}\right)/F_{1}\left(\bar{\mu}\right)$ with $F_{3+m}\left(\bar{\mu}\right)/F_{1}\left(\bar{\mu}\right)\approx  270\zeta(5)/\pi^{2}$ and $7\pi^{2}/10$, respectively for the DA and PA phonons. 
Then, for the characteristic electron temperature relaxation time, we have
\begin{eqnarray}\label{eelrB}
\tilde{\tau}^{s}(T_{e})= \tilde{B}_{s} \left( {\e_0  \over T_{e} } \right)^{1+m} {\bar \tau}_{s}
\end{eqnarray}
where $\tilde{B}_{s}=F_{2}\left(\bar{\mu}\right)/B_{s}F_{1}\left(\bar{\mu}\right)$ with $F_{2}\left(\bar{\mu}\right)/F_{1}\left(\bar{\mu}\right)\approx 18\zeta(3)/\pi^{2}$. The comparison of the respective relaxation rates shows that the PA phonons dominate at $T_{e}<4.3$ K. In graphene, even at low $T_{e}$, quasi-elastic scattering events with the typical phonon energy $a_{s}T_{e}$ ($a_{PA}\ll a_{DA} \ll 1$) dominate the relaxation in the Boltzmann gas. Therefore, the PA phonon relaxation is a slow process at such low temperatures.

\begin{figure}[t]
\includegraphics[width=\columnwidth]{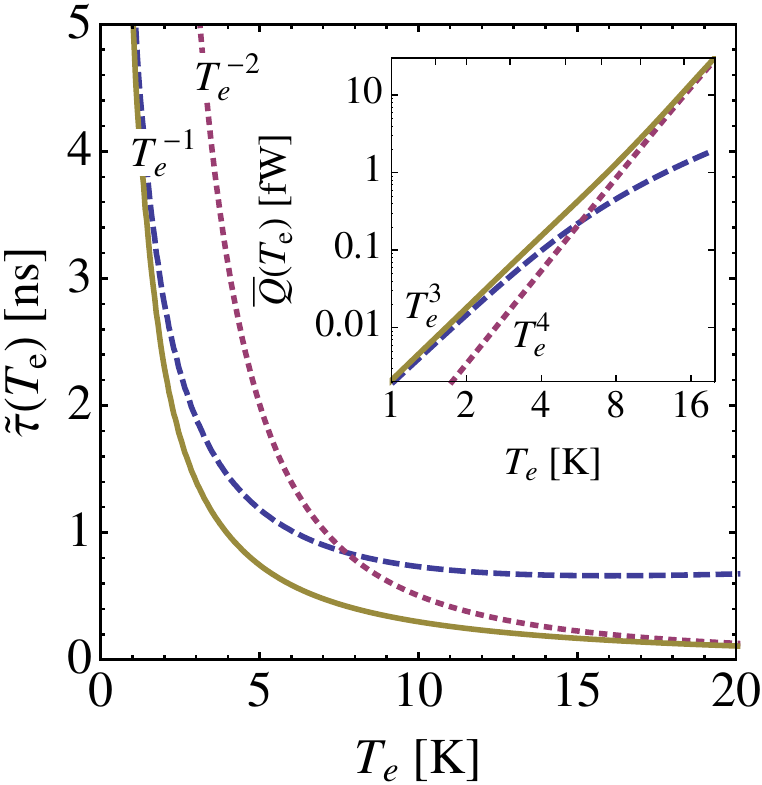}
\caption{Suppression of the DA phonon relaxation at low temperatures in comparison with that through the PA phonons. 
The dotted, dashed, and solid curves plot the characteristic electron temperature relaxation time for the DA, PA, and  PA+DA combined relaxation mechanisms as a function of $T_{e}$ in a Fermi gas with density $n=5\times 10^{13}$ cm$^{-2}$. The inset shows the respective curves for the cooling power $\bar{Q}(T_{e})$.}
\label{fig2}
\end{figure}

\subsection{Electron temperature relaxation in a Fermi gas of Dirac fermions}

This contrasts the much faster relaxation through the PA phonons, we find in a Fermi gas of Dirac fermions ($ T_{e}\ll \e_{F}$). Here the chemical potential is approximated as $\mu(T_{e})\approx \e_{F}-\pi^{2} T_{e}^{2}/6\e_{F}$ and we have $\bar{\e}(T_{e})-\bar{\e}(T)=(\pi^{2}/3)(T_{e}^{2}-T^{2})/\e_{F}$.
In the {\it low temperature BG regime} ($T_{e}\ll T_{BG}^{s}$) the relaxation is dominated by {\it dynamically nonelastic small angle scattering events} with the typical phonon energy $T_{e}$, and for the cooling power in the limit of $T_{e}\gg T$, we find
\begin{eqnarray}\label{cp}
\bar{Q}^s(T_e)&=& A_{s}\left(\frac{\e_{F}}{\e_{0}}\right)^{1+m} \left(\frac{T_{e}}{T_{BG}^{s}}\right)^{2+m} {T_{e}\over {\bar \tau}_{s}}
\end{eqnarray}
where $A_{s}=2 {a^{2}_s c_s^2} (2+m)!\zeta(3+m)/ \pi$. Then, the characteristic temperature relaxation time is
\begin{eqnarray}\label{ETCTf}
\tilde{\tau}^{s}(T_{e})=\tilde{A}_{s} \left(\frac{\e_{0}}{\e_{F}}\right)^{1+m} \left(\frac{T_{BG}^{s}}{T_{e}}\right)^{1+m}{\bar \tau}_{s}
\end{eqnarray}
with $\tilde{A}_{s}=2 \pi^{2} a_{s}/3 A_{s}$. It is seen that in this regime the cooling power shows the same carrier density dependence, $\bar{Q}^s(T_e) \propto 1/\sqrt{n}$, for the PA and DA phonons but the temperature dependence is different, $\bar{Q}^{PA}(T_e)\propto T_{e}^3$ and $\bar{Q}^{DA}(T_e)\propto T_{e}^4$. The crossover from the DA to the PA phonon dominated temperature relaxation of Dirac fermions occurs at $T_{e}\approx 8.9$ K with the characteristic nanosecond relaxation time, fully assessable in experiment. 
In Fig.~\ref{fig2} our numerical calculations from Eqs. (\ref{ETP}), (\ref{ETCT}) without further approximations confirm this analytical prediction of the DA phonon suppression in a Fermi gas of Dirac fermions at low temperatures.

\section{Screening effects}

In this section we discuss the effect of screening by carriers on the PA and DA phonon scattering and on the results obtained in previous sections of the dominance of the PA phonon relaxation mechanism. We take into account the screening effect within the random phase approximation \cite{Schut2011,GV}, which implies the following substitution for the matrix elements
\begin{eqnarray}\label{MES}
M^{\pm s {\bf q}}_{\lambda {\bf k}\rightarrow \lambda' {\bf k}'} \rightarrow \widetilde{M}^{\pm s {\bf q}}_{\lambda {\bf k}\rightarrow \lambda' {\bf k}'}= \frac{M^{\pm s {\bf q}}_{\lambda {\bf k}\rightarrow \lambda' {\bf k}'}}{\e_{scr}(q,\o)}~.
\end{eqnarray}
The screening function is defined as $\e_{scr}(q,\o)=1+ V_{q}\Pi(q,\o)$ where $V_{q}=2\pi e^{2}/(\kappa_{\text{eff}} q)$ is the Fourier transform of the bare Coulomb interaction potential in two-dimensions with the graphene effective dielectric constant $\kappa_{\text {eff}}=(1+\kappa_{\text {GaAs}})/2$ \cite{Landau2,SMB2012} and $\Pi(q,\o)$ is the finite temperature polarization function of Dirac fermions in graphene. In a Fermi gas of Dirac fermions, $\e_{F}\gg T$,
the dynamically nonelastic small angle scattering events with $q\sim T/v_{s}$ and $\o_{s \bf q}\sim T$ dominate the electron relaxation and we have $v_{F} q/2T\gg \o/2T \sim 1$.  Therefore, in this regime for the polarization function we can use the approximation given by Eq.~(43) in the domain IV from Ref.~\onlinecite{Schut2011} and take $\Pi(q,\o)\approx q/4 v_{F}$. Then, the static screening in this regime becomes independent of the phonon wave vector and in GaAs supported graphene, we find $\e_{\text{scr}}\approx 1.5$. Thus, the screening at low temperatures, $T,T_{e}\ll T_{BG}^{s}$, in a Fermi gas does not change the temperature dependence of the relaxation \cite{Efetov2010} and affects {\it similarly} both the PA and DA electron-phonon interaction strengths, reducing them approximately by a factor of $0.67$.

In a Boltzmann gas quasi-elastic scattering events with the much smaller typical phonon energy, $\o_{s \bf q}\sim a_{s}T\ll T$, dominate the electron relaxation. For the polarization function this corresponds to the border between the domains III and IV in Eq.~43 of Ref.~\onlinecite{Schut2011} and needs more careful consideration. From general considerations, however, one can expect that the effect of screening again should be almost similar for the PA and DA scattering mechanisms. This is because in contrast to the electron interaction with bulk piezoelectric acoustical phonons, the PA interaction vertex for piezoelectric surface phonons in Eq.~\ref{PA} has no singularity in the long wavelength limit \cite{Levinson1996}, which would be smeared by screening thereby decreasing strongly the PA electron phonon coupling. 

Thus, we conclude that the results obtained in the previous sections on the regimes, where the PA phonon relaxation mechanism dominates the DA mechanism in the absence of screening, remain true after including the effect of screening.

\section{Summary}

The energy and temperature dependence of the calculated relaxation rates in the different characteristic regimes are summarized in Table~\ref{tab} as being the main results of our paper.
In the Boltzmann gas the temperature dependence of the cooling power per electron, $\bar{Q}^{s}(T_e)\propto T_{e}^{2+m}$, and of the average electron energy, $\bar{\e}^{s}(T_{e})\propto T_{e}$, are weaker than in the Fermi gas, $\bar{Q}^{s}(T_e)\propto T_{e}^{3+m}$ and $\bar{\e}^{s}(T_{e})\propto T^{2}_{e}$. We find that the behavior of the characteristic relaxation time $\tilde{\tau}^{s}(T_{e})$ versus $T_{e}$ is independent of the degree of degeneracy of the electron gas. However, its temperature dependence is different for the PA and DA phonons, i.e. $\tilde{\tau}^{DA}(T_{e})\propto T_{e}^{-2}$ and $\tilde{\tau}^{PA}(T_{e}) \propto T_{e}^{-1}$. Therefore, at low temperatures the extrinsic PA phonons always provide a faster cooling channel for Dirac fermions. Our calculations show also that the PA mechanism for electron energy and temperature relaxation changes qualitatively the relaxation character both in a (non-)degenerate gas of Dirac fermions. These predictions emphasize the effect of the substrate on the relaxation properties of Dirac fermions at low temperatures and create an interesting possibility to study graphene by probing  piezoelectric surface acoustical phonons.

\section{Acknowledgements}

This work was supported by the Flemish Science Foundation (FWO-Vl) and the Methusalem program of the Flemish government.

\end{document}